\documentclass[onecolumn,floatfix,superscriptaddress,showpacs,showkeys,nofootinbib]{revtex4}%
\usepackage{epsfig}
\usepackage{amssymb,latexsym,amsmath}
\newcommand{\eq}[1]{\begin{align} #1 \end{align}}

\usepackage{bm}
\usepackage{amsmath}
\usepackage{graphicx}
\usepackage{subfigure}
\usepackage[colorlinks=true,linktocpage=true,allcolors=blue]{hyperref}
\usepackage[usenames,dvipsnames]{color}

\begin{document}

\title{Fluctuations as a test of chemical non-equilibrium at the LHC}

\author{Viktor Begun}
 \email{viktor.begun@gmail.com}
 \affiliation{Institute of Physics, Jan Kochanowski University, PL-25406~Kielce, Poland}


\begin{abstract}
It is shown that large chemical potential leads to the significant
increase of multiplicity fluctuations for bosons, and makes the
fluctuations infinite in the case of Bose-Einstein condensation.
It allows to distinguish between the models that explain the
anomalous proton to pion ratio and the low transverse momentum
enhancement of pion spectra in Pb+Pb collisions at the LHC within
chemical equilibrium or non-equilibrium models.
The effects of resonance decays, finite size of the system,
requirements to the event statistics, different momentum cuts, and
limited detector acceptance are considered.
The obtained results show the possibility to observe a substantial
increase of the normalized kurtosis for positively or negatively
charged pions in the case of non-equilibrium or partial pion
condensation using currently measured data.
\end{abstract}



\maketitle

\section{Introduction}
\label{sect:Intro}

During last decades thermal
model~\cite{Fermi:1950jd,Hagedorn:1965st,Cleymans:1992zc,Sollfrank:1993wn,Schnedermann:1993ws,BraunMunzinger:1994xr,Becattini:2000jw,Florkowski:2001fp,Broniowski:2001we,BraunMunzinger:2003zd}
(TM) became a standard tool for the analysis of mean
multiplicities in nucleus-nucleus collisions. It is implemented in
free online
codes~\cite{Wheaton:2004qb,Torrieri:2004zz,Petran:2013dva}, and
obtained temperatures are discussed as a basic property of a
created system in the papers reporting experimental results, see
e.g.~\cite{Abelev:2013vea,Floris:2014pta}.
The temperatures follow a smooth freeze-out line in a wide energy
range of colliding nuclei~\cite{Cleymans:2005xv,Andronic:2005yp}.
%
%
The initial energy of the collision is 10 times larger at the LHC
(Large Hadron Collider) than at RHIC (Relativistic Heavy Ion
Collider) and 100 times larger than at SPS (Super Proton
Synchrotron).
The temperature grows with increasing energy of the collision, and
was expected to saturate around $T\simeq165~$MeV.
Therefore, the LHC data~\cite{Abelev:2012wca,Abelev:2013vea} came
as a big surprise, because their description requires that the
temperature falls down from the freeze-out line by
$10$~MeV~\cite{Floris:2014pta,Stachel:2013zma}, being smaller than
at RHIC and close to that at the SPS\footnote{The recent analysis
of the new SPS data gives a different freeze-out line, which
points to the low LHC temperature~\cite{Vovchenko:2015idt}. The
same TM used for the LHC confirms this
finding~\cite{Vovchenko:2015cbk}.}.
This difference in temperature is very large for a TM, because all
particles except for pions have the mass $m\gg165$~MeV, and their
mean multiplicities in TM depend exponentially on temperature
$\langle N\rangle\sim\exp[-m/T]$.

Besides the lower temperature at the LHC, there are substantial
difficulties in simultaneous description of pions, protons and
strange particles. Proton to pion ratios are suppressed at the LHC
compared to RHIC~\cite{Abelev:2013vea}. Experimentally measured
pion spectra at the LHC rise steeper for low transverse momentum
$p_T$ than in the
models~\cite{Molnar:2014zha,Ryu:2015vwa,Naboka:2015qra}, while the
same models work perfectly at RHIC.
There are many ways to explain the proton to pion ratio or strange
particles~\cite{Becattini:2012xb,Noronha-Hostler:2014aia,Chatterjee:2013yga,Chatterjee:2014lfa,Naskret:2015pna,Prorok:2015vxa},
but the low $p_T$ enhancement of the pion spectrum at the LHC, is
still the open problem.

Both, proton to pion ratio and the low $p_T$ pion spectrum can be
explained in the non-equilibrium
TM~\cite{Petran:2013lja,Begun:2013nga}. It allows for a
non-equilibrium chemical potential\footnote{The non-equilibrium
can describe proton spectra as good as the
rescattering~\cite{Ryu:2015vwa}, but only a non-equilibrium
chemical potential can give the low $p_T$ enhancement of pions
seen in the data~\cite{Begun:2013nga}.} for each particle, due to
partial equilibration of the constituent quarks in the fast
expanding fireball~\cite{Koch:1985hk,Rafelski:2015cxa}. This model
has two more parameters compared to the standard TM - one for
light and one for strange quarks. The numerical calculations in
the non-equilibrium TM give even smaller temperature
$T\simeq140$~MeV, and the large positive chemical potential for
pions close to it's mass $\mu_{\pi}\simeq
m_{\pi}$~\cite{Petran:2013lja}, see also~\cite{Melo:2015wpa}. This
may imply the Bose-Einstein condensation (BEC) of
pions~\cite{Begun:2014aha,Begun:2015ifa,Begun:2015yco}.

The pion chemical potential was introduced to explain the early
data at the SPS~\cite{Kataja:1990tp}, and was similarly justified
by partial thermalization \cite{Gavin:1991ki}, and also by pion
condensation \cite{Gerber:1990yb,Ornik:1993gb,Turko:1993dy}.
However, the update of the resonance list gave the same
effect~\cite{Schnedermann:1993ws}, and pion BEC was abandoned.
It seems not to be the case at the LHC, because the properties of
the resonances with $m<2.5$~GeV are known very well now. They are
already included in TM, and do not give the required amount of low
$p_T$ pions. The resonances with $m>2.5$~GeV may give the effect,
if the particular type of Hagedorn-like states that decay mainly
into low $p_T$ pions exist. The deficiency of pions at the LHC is
observed at the $p_T\leq150$~MeV~\cite{Begun:2015ifa}. It means
that the not yet observed Hagedorn-like states with $m>2.5$~GeV
should decay through multi-pion channels with 5-10 pions, or
through a particular sequential decay, that gives many low $p_T$
pions. The only possible light meson candidate was the famous
sigma $f_0(500)$ meson, but it should {\it not} be included in TM
at
all~\cite{Broniowski:2015oha,GomezNicola:2012uc,Pelaez:2015qba,Giacosa:2016rjk}.

There are good reasons for chemical non-equilibrium with
$\mu_{\pi}>0$ at the LHC. It was predicted in the super(over)
cooling scenario~\cite{Csorgo:1994dd,Shuryak:2014zxa}. The extra
pions at low $p_T$ may appear due to fast hadronization of the
gluon condensate~\cite{Blaizot:2011xf,Blaizot:2012qd},
glueballs~\cite{Vovchenko:2015yia}, or Color Glass
Condensate~\cite{Iancu:2000hn,Iancu:2003xm}, forming transient
Bose-Einstein condensate of pions~\cite{Gelis:2014tda}.
The time needed to form such a condensate at the LHC is lower than
at RHIC, and is just
$t\sim0.1-0.2$~fm/c~\cite{Scardina:2014gxa,Meistrenko:2015mda}.
The analysis of two-, three- and four-particle correlations by
ALICE Collaboration~\cite{Abelev:2013pqa,Adam:2015pbc} gives large
values for the amount of pions from a coherent source - $20-30\%$.
They do not specify the nature of the coherent emission, but pion
condensate is a good candidate.

A large positive chemical potential substantially increases
multiplicity fluctuations of bosons and makes the fluctuations
{\it infinite} for the case of pion BEC in an infinite
system~\cite{Begun:2005ah,Begun:2006gj}. It may allow to differ
pion BEC from other effects like the production of Hagedorn-like
states discussed above, or the disoriented (disordered) chiral
condensate (DCC), see e.g.~\cite{Blaizot:1992at,Mohanty:2005mv}.
At the LHC the radius of the system at freeze-out is
$r\sim10$~fm~\cite{Begun:2014rsa}, and the amount of pions on the
zero momentum (condensate) level might be around
$5\%$~\cite{Begun:2015ifa}.
However, it can be enough to observe a detectable signal in pion
multiplicity fluctuations\footnote{The high order fluctuations
received a lot of attention recently due to a possibility to
detect QCD critical point, see
e.g.~\cite{Schuster:2009jv,BraunMunzinger:2011ta,Bzdak:2012an,Bhattacharyya:2013oya,Alba:2014eba,Nahrgang:2014fza,Asakawa:2015ybt,Gupta1525,Aggarwal:2010wy},
however, it seems that pion fluctuations with $\mu_{\pi}\gg0$ were
not studied yet.}, using currently measured events by ALICE. If
the fluctuations will be found small, then it will be a strong
argument against non-equilibrium at the LHC. However, if the
fluctuations will be found large, then one could use them as a
tool to study the non-equilibrium.

The paper is organized as follows. In Section~\ref{sect:Prim} the
phase diagram of the pion gas is obtained in order to determine
the centrality where the largest amount of the condensate is
possible at the LHC. In Section~\ref{sect:Fluc} the fluctuations
of primary pions are calculated, and further suggestions how to
search for the condensate are formulated. In
Section~\ref{sect:Res} the resonance decay contribution,
requirements to event statistics, and the effects of limited
detector acceptance are considered. A specific $p_T$ cut is
proposed to enhance the effect of possible pion condensation.
Section~\ref{sect:Concl} concludes the paper.
%

\section{Phase diagram of the condensate}
\label{sect:Prim}

Bose-Einstein condensation is possible at any temperature, if the
density of bosons $\rho$ is high enough
 \eq{
 \rho(T,\mu) ~=~ \frac{V}{(2\pi)^3}\int ~\frac{d^3p}{\exp\left[(E_p-\mu)/T\right]-1}~,
 \label{Rho}
 }
where $V$ is the system volume, and $E_p=\sqrt{p^2+m^2}$ is the
energy of a boson with a momentum $p$.
The critical density is defined in TM as
$\rho_C(T)=\rho(T,\mu=m)$, that gives a continuous condensation
line in the $T-\rho$ plane~\cite{Begun:2006gj,Begun:2008hq}.
Therefore, one can also find the condensation temperature $T_C$
for each density $T_C(\rho)$.

Multiplicity fluctuations rise to infinity at the condensation
line in the infinite volume limit~\cite{Begun:2006gj}, and
increase fast in it's vicinity for a finite volume of the
system~\cite{Begun:2008hq}. Therefore it is important to know how
far the system is from the condensation line.
The finite volume corrections were implemented
in~\cite{Begun:2014aha} to SHARE model~\cite{Petran:2013dva}.
The corresponding fit of the mid-rapidity yields
$\frac{dN_i}{dy}|_{|y|<0.5}$ at the LHC confirms that chemical
potential is relevant only for pions, giving a smaller value than
in~\cite{Petran:2013lja}.
In this model chemical potential is the same for charged and
neutral pions, so neutral pions could 'feel' the condensation
effects at smaller $\mu_{\pi}$, due to lower mass. However, their
multiplicity is not measured yet, and the spectrum is available
only for $p_T>700$~MeV~\cite{Abelev:2014ypa}, while any effect of
the condensate on spectra can be seen for much smaller momenta
$p_T<200$~MeV~\cite{Begun:2015ifa}. Moreover, in order to address
fluctuations, the number of particles should be measured
event-by-event, which is even more complicated.
The number of positively and negatively charged pions is the same
within the error bars~\cite{Abelev:2013vea}, therefore
$\mu_{\pi^+}\simeq\mu_{\pi^-}\equiv\mu$ and there is no difference
which one to use. However, charge identification is important,
because $\pi^+$ and $\pi^-$ are different particles that condense
separately.

\begin{figure}[h!]
 \includegraphics[width=0.49\textwidth]{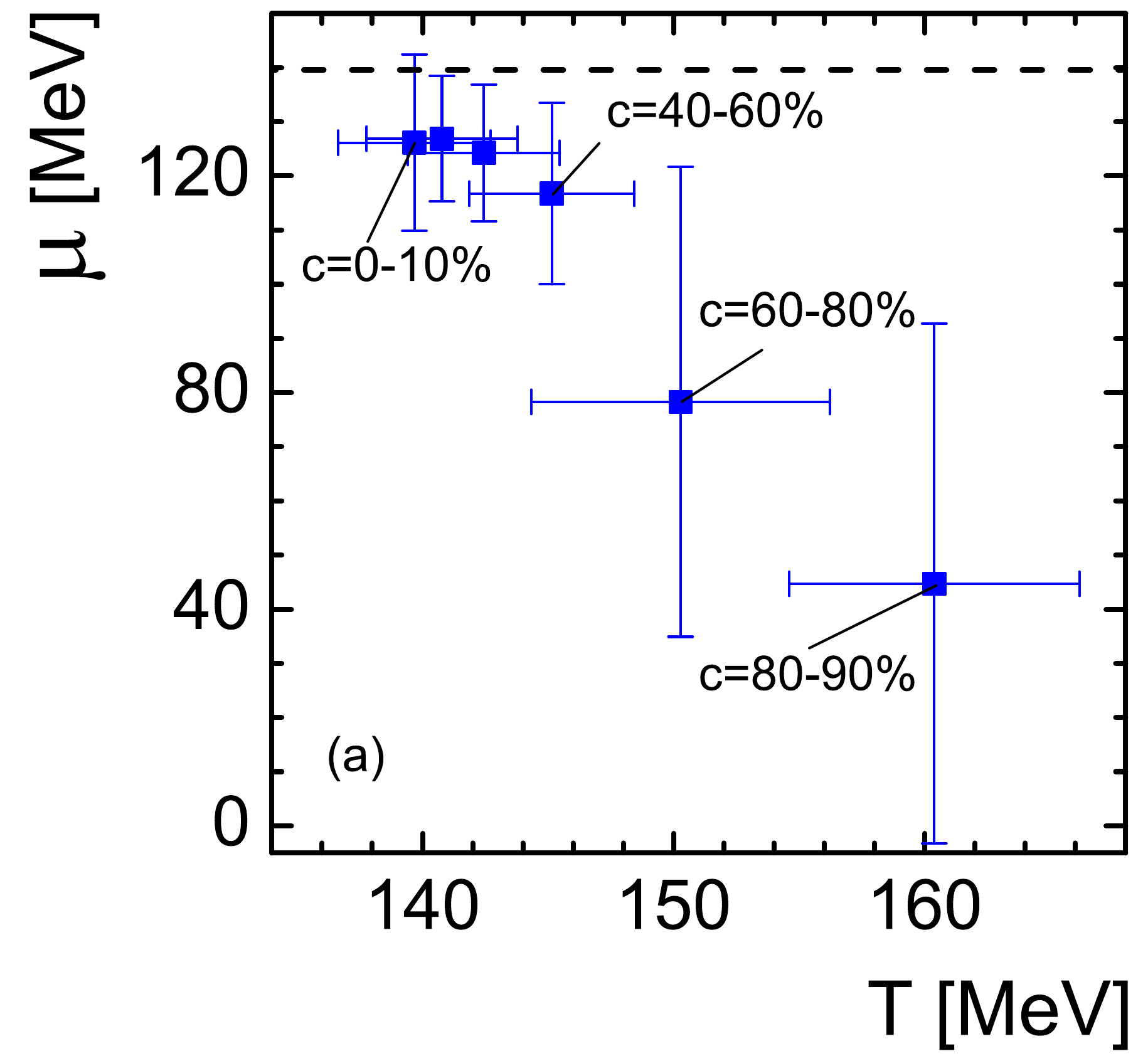}~~
 \includegraphics[width=0.49\textwidth]{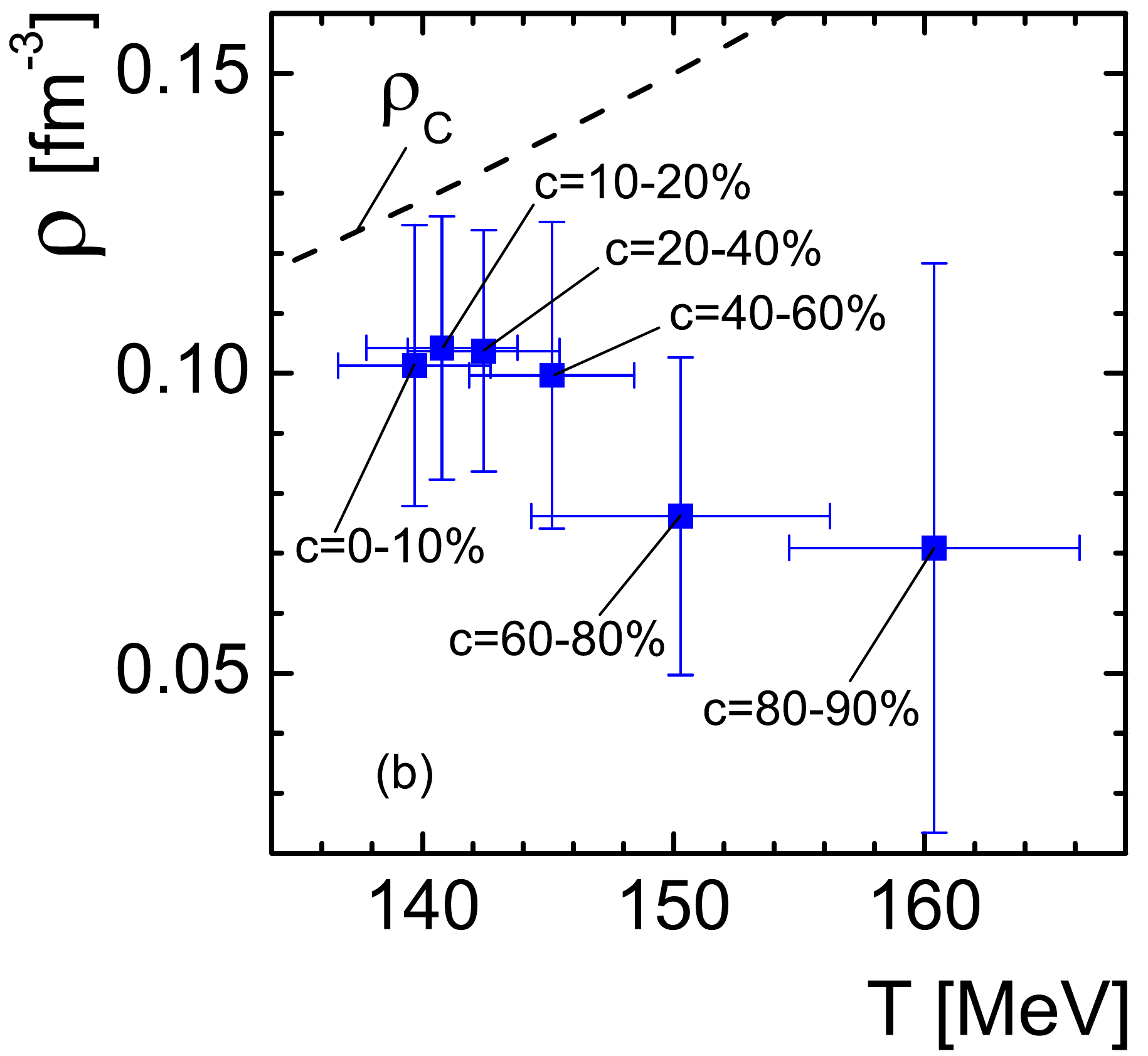}
 \caption{(a): The pion chemical potential and temperature at freeze-out for different centralities in the $2.76$~TeV Pb+Pb collisions at the
 LHC in the non-equilibrium TM~\cite{Begun:2014aha}. (b): The same for densities and temperatures.
 The dashed lines show the chemical potential $\mu=m_{\pi^{\pm}}$ and the critical density $\rho_C(T)$.}\label{fig:T-mu}
\end{figure}

The densities and chemical potentials for positively or negatively
charged pions are calculated for different centralities of the
collision at the LHC, using the parameters obtained
in~\cite{Begun:2014aha}, and are shown in Fig.~\ref{fig:T-mu}.
One can see that $\rho<\rho_C$ and $T>T_C$ at the LHC, so the
condensate line is not reached, but central and semi-central
collisions with centrality $c<40\%$ are the closest to the
condensation line. Note a small temperature for the most central
collisions $T\simeq140$~MeV as
in~\cite{Petran:2013lja,Petran:2013qla}, which increases for
peripheral collisions and reaches the equilibrium TM result of
$T\simeq150-160$~MeV~\cite{Stachel:2013zma,Vovchenko:2015idt} for
very peripheral collisions. The chemical potential decreases for
peripheral collisions in contrast to~\cite{Petran:2013lja},
because of the finite size of the system at freeze-out. Therefore,
the condensation is more probable in the most central collisions,
where the system is also spatially larger and lives longer.

The error bars are obtained using the standard methods for
propagation of uncertainty. The necessary correlations of the
parameters are calculated for the $10\%$ deviation of the
$\chi^2/N_{\rm dof}$ from the best fit~\cite{Begun:2015ifa}. The
correlation between all pairs of thermodynamic parameters is
negative at all centralities, except for a small positive
correlation between $V$ and $T$ at $c=60-80\%$, and between $T$
and $\mu_{\pi}$ at $c=80-90\%$. Therefore, the error bars are the
largest at these centralities. However, they are significant also
at other centralities. It reflects the freedom in choosing the
parameters to fit the available data.
A larger set of measured mean multiplicities should decrease this
ambiguity.

\section{Fluctuations of primary pions}\label{sect:Fluc}

Multiplicity fluctuations of any order can be calculated for
primary pions analytically, using the definition of
susceptibilities $\chi_n$. They are given by the derivatives of
pressure ${\cal P}$ by chemical potential $\mu$ at constant
temperature $T$, see e.g.~\cite{Karsch:2010ck,Asakawa:2015ybt}
 \eq{\label{Chi-n}
 \chi_n = \frac{\partial^n({\cal P}/T^4)}{\partial (\mu/T)^n}\Big|_{T}~.
 }
The pressure in the pion gas is given by
 \eq{\label{PT4}
 {\cal P}/T^4 ~=~ \frac{1}{T^3}\,\sum_p
 \ln(1-\exp[(\mu-E_p)/T])^{-1}~,
 }
%
The convenient measures are the scaled variance (variance over the
mean) $\omega = \sigma^2/\langle N\rangle$, normalized skewness
$S\cdot\sigma$, and normalized kurtosis\footnote{Note that for
Gauss (Normal) distribution $\omega$ can get any value, while
$S\cdot\sigma=\kappa\cdot\sigma^2\equiv0$.} $\kappa
\cdot\sigma^2$.
They are directly related to the susceptibilities and central
moments
 \eq{\label{Fluc}
 \omega ~=~ \frac{\chi_2}{\chi_1} ~=~ \frac{m_2}{\langle N\rangle}~,
 &&
 S\cdot\sigma ~=~ \frac{\chi_3}{\chi_2} ~=~ \frac{m_3}{m_2}~,
 &&
 \kappa\cdot\sigma^2 ~=~ \frac{\chi_4}{\chi_2} ~=~ \frac{m_4}{m_2} ~-~3\,m_2~,
 }
where $\langle N\rangle$ is the mean multiplicity and
 \eq{\label{m_n}
 m_n~=~\langle(N-\langle N\rangle)^n\rangle~=~\sum_N(N-\langle N\rangle)^n\cdot P(N)
 }
are the central moments of the $P(N)$ multiplicity distribution.
Equation (\ref{Chi-n}) is very useful for theoretical
calculations, while Eq.~(\ref{m_n}) is better for
experimentalists, because they directly measure the $P(N)$.
The straightforward calculation using
Eqs.~(\ref{Chi-n})-(\ref{Fluc}) give:
 \eq{\label{Np}
 \langle N\rangle
 &~=~ \sum_p \langle n_p \rangle~,
 \\ \label{w}
 \omega
 &~=~ \frac{\sum_p \left( \langle n_p \rangle ~+~  \langle n_p \rangle^2\right)}{\sum_p \langle n_p \rangle}
  ~=~ 1 ~+~ \frac{\sum_p \langle n_p \rangle^2}{\sum_p \langle n_p \rangle}~,
 \\
 S\cdot\sigma
 &~=~ \frac{\sum_p \left( \langle n_p \rangle ~+~ 3\langle n_p \rangle^2 ~+~  2\langle n_p\rangle^3\right)}
           {\sum_p \left( \langle n_p \rangle ~+~  \langle n_p \rangle^2\right)}~,
 \\
 \kappa\cdot\sigma^2
 &~=~ \frac{\sum_p \left( \langle n_p \rangle   ~+~ 7\langle n_p \rangle^2
                    ~+~ 12\langle n_p \rangle^3 ~+~ 6\langle n_p \rangle^4 \right)}
           {\sum_p \left( \langle n_p \rangle   ~+~  \langle n_p \rangle^2\right)}~,
 \label{ks2p}
 }
where
%
$ \langle n_p \rangle^k ~=~
 \left\{\exp\left[(\sqrt{{p}^2+m^2}-\mu)/T\right]-1\right\}^{-k}$.

In equilibrium $\mu=0$, and Eqs.~(\ref{w})-(\ref{ks2p}) give for
positively (negatively) charged pions at $T=140$~MeV:
 \eq{\label{fluc-Pi}
 \omega~\simeq~1.1~, && S\cdot\sigma~\simeq~1.2~, && \kappa\cdot\sigma^2~\simeq~1.9~.
 }
For $\mu=0$ and $m/T\rightarrow\infty$ one recovers the result for
Boltzmann statistics with
$\omega=S\cdot\sigma=\kappa\cdot\sigma^2=1$. The $\mu=0$ and
$m/T\rightarrow0$ is never realized in TM, because the
temperatures are usually of the order of the pion mass or lower.
However, one can see that for this case the scaled variance is
finite, $\omega\simeq1.368$~\cite{Begun:2005ah}, but
$S\cdot\sigma$ and $\kappa\cdot\sigma^2$ diverge on the lower
bound of the momentum integral, that usually replaces the sum over
the momentum levels in (\ref{Np})-(\ref{ks2p}) $\sum_p \rightarrow
V/(2\pi^2)\int p^2dp$.
For $\mu\rightarrow m$ even $\omega$ diverges, as well as all
$\int p^2dp\,\langle n_p \rangle^k$ with $k\geq2$. This is the
consequence of the fact that Bose-Einstein condensation is the 3rd
order phase transition\footnote{The similar divergences in high
order fluctuations measures take place close to critical
point~\cite{Stephanov:2008qz}.}.
However, there are no divergences in finite volume, because the
maximal fluctuations are bounded by the number of particles in the
system. One can take the finite volume into account keeping the
zero momentum state in the sum $\sum_p$
 \eq{\label{sum-int}
 \sum_p\langle n_p\rangle^k ~~\longrightarrow~~ \langle n_0\rangle^k ~+~ \frac{V}{2\pi^2}\int_0^{\infty} \langle n_p\rangle^k~p^2dp~,
 }
because $\langle n_0\rangle=\frac{1}{\exp[(m-\mu)/T]-1}$ grows as
fast as volume in the limit $\mu\rightarrow
m$~\cite{Begun:2008hq}. The corresponding competition between
$\mu$ and $V$ during the fit of pion mean multiplicities led to
the decrease of pion chemical potential
in~\cite{Begun:2014aha,Begun:2015ifa} compared
to~\cite{Petran:2013lja}.
The relative contribution of the first term on the right hand side
of Eq.~(\ref{sum-int}) is larger for $\mu\rightarrow m$, because
the largest contribution to the integral comes from the lower
bound $p\rightarrow0$, which diverges as $\langle n_0\rangle^k$ in
this limit, but the $p^2dp\rightarrow0$ weakens the divergency.
Therefore, at $\mu\rightarrow m$ one can estimate the fluctuations
assuming that there is only the condensate level $p=0$. Keeping
also only the highest $k$ in Eqs.~(\ref{w})-(\ref{ks2p}) gives
 \eq{\label{wd}
 \omega ~\simeq~ \frac{1}{\langle N\rangle}\,\frac{1}{\delta^2}~,
 &&
 S\cdot\sigma ~\simeq~ \frac{2}{\delta}~,
 &&
 \kappa\cdot\sigma^2 ~\simeq~ \frac{6}{\delta^2}
 }
where $\delta=(m-\mu)/T$. The approximation (\ref{wd}) is valid,
if $\langle n_0\rangle^2\gg\langle N\rangle$,  i.e. for
$\omega\gg1$.
Let us also assume for simplicity that the condensation line is
already reached, because it excludes the $\rho-\rho_C$ dependance.
Then, $\delta=(aV)^{-2/3}$, where
$a=(mT)^{3/2}/(\sqrt{2}\pi)$~\cite{Begun:2008hq} and one obtains
 \eq{
 \omega
 ~\simeq~ \frac{a}{\rho_C}\,(a\,V)^{1/3},&&
 S\cdot\sigma
 ~\simeq~ 2\,(a\,V)^{2/3} ~\sim~ \omega^2,&&
 \kappa\cdot\sigma^2
 ~\simeq~ 6\,(a\,V)^{4/3} ~\sim~ \omega^4~.
 \label{wV}
 }
Therefore, the higher is the order of fluctuations, the faster
they grow.

Equations~(\ref{sum-int},\ref{wd}) suggest that the fluctuations
at $\mu\rightarrow m$ should increase, if one finds a way to
increase the relative amount of registered particles on the $p=0$
level, see~\cite{Begun:2016egl}. It can be done by applying the
$p_T$ cut that selects more pions from the condensate $\langle
n_0\rangle$.
The pion spectra at the LHC are measured starting from
$p_T>100$~MeV. Pions on the $p=0$ level can receive a momentum
$p_T\lesssim 200$~MeV, because of the collective motion with the
hypersurface~\cite{Begun:2015ifa}. Therefore, three distinct cases
can be considered:
 \begin{itemize}
 \item{\makebox[4cm][l]{all $p_T$}         -~~the easiest to calculate, but hard to measure,}
 \item{\makebox[4cm][l]{$p_T>100$~MeV}     -~~currently measured data,}
 \item{\makebox[4cm][l]{$p_T=100-200$~MeV} -~~contais the highest percentage of pions from the $p=0$ level.}
 \end{itemize}

Fluctuations of primary pions, both normal and those from the
condensate, can be calculated in Cracow single freeze-out
model~\cite{Broniowski:2001uk,Baran:2003nm,Kisiel:2006is,Chojnacki:2011hb}.
It can be done taking numerically the integral over the
hypersurface, and for the corresponding $p_T$ intervals $\Delta
p_T^{\rm norm}$, similar to the case with just the spectra in
Ref.~\cite{Begun:2015ifa}
 \eq{\label{np}
 \sum_p \langle n_p\rangle^k
 &~=~ \frac{1}{\{\exp[(m-\mu)/T]-1\}^k}\, \frac{\Delta p_{T}^{\rm cond}}{p_T^{\rm max}}
 \\
 &~+~ \frac{1}{(2\pi)^3}\int_{\Delta p_T^{\rm norm}} p_Tdp_T
   \int_0^{2\pi}  d\phi_p
   \int_0^{2\pi}  d\phi
   \int_{-\infty}^{\infty} d\eta_{||}
   \int_0^{r_{\rm max}} rdr
 \nonumber\\
 &\hspace{0.5cm}~\times~ \left[m_T\,\sqrt{\tau_f^2+r^2}\cosh(\eta_{||}-y)-p_Tr\cos(\phi-\phi_p)\right]
 \nonumber\\
 &\hspace{0.5cm}~\times~
 \left\{\exp\left(\frac{1}{T}\left[m_T\sqrt{1+\frac{r^2}{\tau_f^2}}\cosh(\eta_{||}-y)-p_T\,\frac{r}{\tau_f}\cos(\phi-\phi_p)\right]-\frac{\mu}{T}\right)-1\right\}^{-k}~,
 \nonumber
 }
where the first term is the contribution from the condensate, and
$\Delta p_{T}^{\rm cond}$ is the interval where the corresponding
$p_T$ cut overlaps with the condensate. The maximal momentum of
the condensate, $p_T^{\rm max}=m\,r_{\rm max}/\tau_f$, is
determined by the radius of the hypersurface $r_{\rm max}$ and the
freeze-out time $\tau_f$~\cite{Begun:2015ifa};
$m_T=\sqrt{m^2+p_T^2}$ is the transverse mass, and $ \int_0^{2\pi}
d\phi \int_{-\infty}^{\infty} d\eta_{||}\int_0^{r_{\rm max}} rdr$
is the integration over the hypersurface. The integral over all
$p_T$ gives the same as the integral over the volume per unit
rapidity $V=\pi r_{max}^2\tau_f$~\cite{Begun:2014rsa}. The
integral over rapidity $dy$ is absent in the right hand side of
Eq.~(\ref{np}), because the fit of thermodynamic parameters was
done for the rapidity densities
$\frac{dN_i}{dy}|_{|y|<0.5}$~\cite{Begun:2014aha,Begun:2015ifa}.

Another possibility to enhance the fluctuations is the increase of
volume\footnote{The rapidity distributions are flat in the wide
range of rapidities at the LHC. Thus, temperature and chemical
potentials should not change, while total volume increases with
increasing the rapidity interval.}, as seen from Eq.~(\ref{wV}).
It can be done by increasing the rapidity interval where pions are
measured.
%
It should be noted that the same assumption as
in~\cite{Begun:2015ifa} is made so far, that the coherence length
of the condensate in rapidity, $\Delta y_{\rm cond}$, is the same
as the rapidity interval of the measurements.
If $\Delta y_{\rm cond}$ is much larger, then one could use the
approximate formula (\ref{acc}) from the next section.
If $\Delta y_{\rm cond}$ is smaller and fluctuations of the
condensate come from the uncorrelated parts of the freeze-out
hypersurface, or just from a small part of it, then the
fluctuations will be smaller and scale differently from
(\ref{wV}).
In any case, the $\kappa\cdot\sigma^2$ observable seems to be
sensitive enough to study these effects.

\section{Resonance Decay Contribution}\label{sect:Res}

The question about fluctuations in a real system can be addressed
semi-analytically under the assumption that the system consists of
two parts that do not correlate.
It seems to be a reasonable approximation, because the
corresponding fluctuations are very different. As we will see, the
fluctuations of pions from resonance decays in small acceptance
window in rapidity are $\sim 1$, as for Poisson distribution.
At the same time, pion fluctuations rapidly increase at
$\mu\rightarrow m$, see Eqs.~(\ref{wd})-(\ref{wV}).
%

For two uncorrelated multiplicity distributions $P_1(N_1)$ and
$P_2(N_2)$ one has:
 \eq{\label{N12}
 \langle N\rangle
 &~=~ \langle N_1\rangle ~+~ \langle N_2\rangle~,
 \\
 \omega
 &~=~ \omega_1~\frac{\langle N_1\rangle}{\langle N\rangle} ~+~ \omega_2~\frac{\langle N_2\rangle}{\langle N\rangle}~,
 \\
 S\cdot\sigma
 &~=~ S_1\cdot\sigma_1~\frac{\omega_1}{\omega}~\frac{\langle N_1\rangle}{\langle N\rangle}
  ~+~ S_2\cdot\sigma_2~\frac{\omega_2}{\omega}~\frac{\langle N_2\rangle}{\langle N\rangle}~,
 \\
 \kappa\cdot\sigma^2
 &~=~ \kappa_1\cdot\sigma_1^2~\frac{\omega_1}{\omega}~\frac{\langle N_1\rangle}{\langle N\rangle}
  ~+~ \kappa_2\cdot\sigma_2^2~\frac{\omega_2}{\omega}~\frac{\langle N_2\rangle}{\langle N\rangle}~.
 \label{ks212}
 }
Therefore, one can calculate primary fluctuations using
Eqs.~(\ref{Np})-(\ref{ks2p}), (\ref{np}) and then mix them with
the fluctuations of pions from resonance decays using
Eqs.~(\ref{N12})-(\ref{ks212}).

The limited detector acceptance can be taken into account similar
to Refs.~\cite{Heiselberg:2000fk,Mahapatra:2001ag,Begun:2004gs}.
In the limit of a very small acceptance window one can neglect all
correlations, use binomial distribution for the probability $q$
for a particle to be accepted, $0\leqslant q\leqslant 1$,
$q\rightarrow0$, and obtain
 \eq{\label{acc}
 \omega
 = 1+q\,(\omega_{\rm all}-1)~,&&
 S\cdot\sigma
 \simeq 1+2q\,(\omega_{\rm all}-1)~,&&
 \kappa\cdot\sigma^2
 \simeq 1+6q\,(\omega_{\rm all}-1)~,
 }
where $\omega_{\rm all}$ is the scaled variance for the case when
all particles are accepted.
One can see from Eq.~(\ref{acc}) that for $\omega_{\rm all}>1$ the
fluctuations of the accepted particles are always larger than
unity and approach to it from above in the small acceptance
limit\footnote{Global conservation of charges, energy and momentum
significantly suppress fluctuations making $\omega_{\rm all}<1$,
see~\cite{Begun:2004gs,Begun:2004pk,Hauer:2007im,Lungwitz:2007uc}.
Therefore, in the case when global conservations start to play a
role the fluctuations approach to unity from
below~\cite{Begun:2006uu}.}. Equation~(\ref{acc}) is an
approximation that should be valid for pions from resonance
decays, but it is not valid if there is some dependance on $p_T$.
For example, the relative amount of primary pions from the
condensate at $p=0$ increases after the application of the cut
with $p_T<200$~MeV, because they are situated only
there~\cite{Begun:2015ifa}. However, Eq.~(\ref{acc}) is still
useful, because it shows that the increase of acceptance leads,
first of all, to the change of higher order fluctuations.

The THERMINATOR model~\cite{Chojnacki:2011hb} is used for the
account of resonances in this paper. Primary particles are sampled
with Poisson distribution there, i.e. $\omega_{\rm prim}=S_{\rm
prim}\cdot\sigma_{\rm prim}=\kappa_{\rm prim}\cdot\sigma_{\rm \rm
prim}^2\equiv1$.  It is not correct for primary pions, because
their number should be sampled according to Bose-Einstein
distribution following Eqs.~(\ref{Np})-(\ref{ks2p}). However, it
gives a good estimate of pion fluctuations due to resonance
decays, because resonances are heavy, and one can use Boltzmann
statistic for them, see the discussion after Eq.~(\ref{fluc-Pi}).
Resonance decays can only increase fluctuations in this case. The
effects of resonance decays are stronger for higher temperatures,
because of the exponential suppression of heavy particles in TM.
The temperature is the highest at $c=80-90\%$ centrality, see
Fig.~(\ref{fig:T-mu}). Therefore, resonance decays at this
centrality give the upper bound on the fluctuations from
resonances at all centralities
 \eq{\label{Fluc-Dec}
 \omega_{\rm res} \lesssim 1.05,&&
 S_{\rm res}\cdot\sigma_{\rm res} \lesssim 1.1,&&
 \kappa_{\rm res}\cdot\sigma_{\rm res}^2 \lesssim 1.3~.
 }
Looking at the numerical values in Eq.~(\ref{Fluc-Dec}) one can
conclude that the scaling (\ref{acc}) holds even quantitatively.
Any $p_T$ cut further decreases the fluctuations for  resonances.
Therefore, the approximation $\omega_{\rm res}=S_{\rm
res}\cdot\sigma_{\rm res}=\kappa_{\rm res}\cdot\sigma_{\rm res}^2
=1$ is used from here on.

Statistical errors increase extremely fast for normalized skewness
and kurtosis when mean multiplicity increases. Using the
definitions for the absolute and relative errors of the unknown
variable $X$
 \eq{
 X=\langle X\rangle \pm \sigma(X),&&
 \varepsilon_X=\frac{X-\langle X\rangle}{\langle X\rangle}~,
 }
one obtains for the mean multiplicity, scaled variance, normalized
skewness and kurtosis~\cite{MathWorld}:
 \eq{
 \varepsilon_{\langle N\rangle} \simeq \frac{1}{\sqrt{N_{\rm ev}}\sqrt{\langle N\rangle}},&&
 \varepsilon_{\omega} \simeq \sqrt{\frac{3}{N_{\rm ev}}},&&
 \varepsilon_{S\cdot\sigma} \simeq \sqrt{\frac{6}{N_{\rm ev}}}\sqrt{\langle N\rangle},&&
 \varepsilon_{\kappa\cdot\sigma^2} \simeq \sqrt{\frac{24}{N_{\rm ev}}}\,\langle N\rangle~,
 }
where $N_{\rm ev}$ is the number of generated events and $\langle
N\rangle$ is the mean multiplicity. Therefore, in order to have a
relative error for the normalized kurtosis on the level of
$\varepsilon_{\kappa\cdot\sigma^2}=10\%$, one has to generate
$N_{\rm ev}=24*10^2*\langle N\rangle^2$ events. For pions in the
most central collisions at the LHC it gives the number $N_{\rm
ev}\sim10^9$.
%
For smaller statistics one can obtain huge and even negative
values for $S\cdot\sigma$ and $\kappa\cdot\sigma^2$ which
fluctuate with $N_{\rm ev}$ just because of small statistics.
\begin{figure}
 \includegraphics[width=0.49\textwidth]{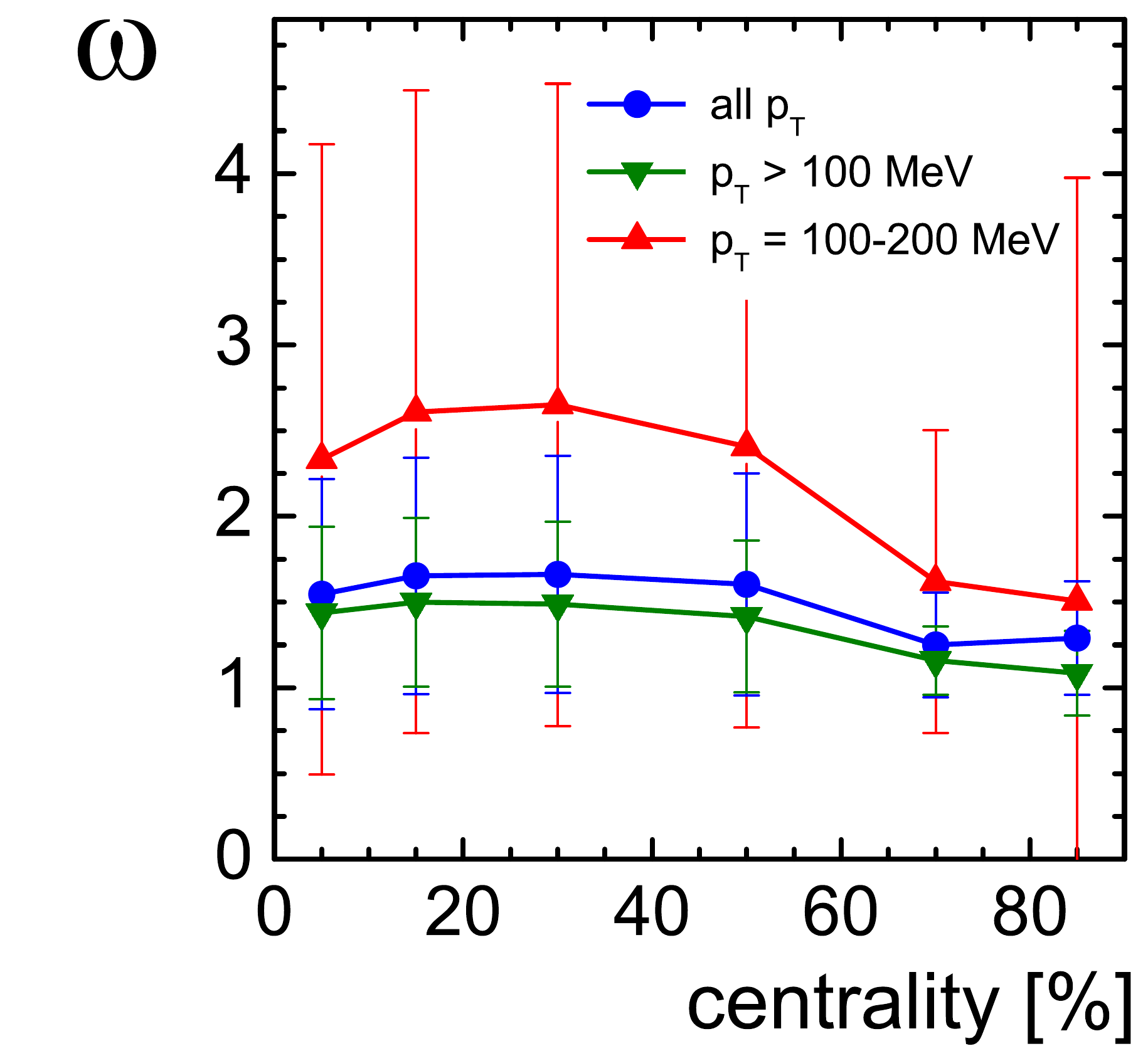}
 \caption{
 The total fluctuations of positively (negatively) charged pions. The resonance decays and the
 condensate for different $p_T$ cuts as the function of the collision centrality are included.
 \label{fig:W}}
\end{figure}
\begin{figure}[h!]
 \includegraphics[width=0.49\textwidth]{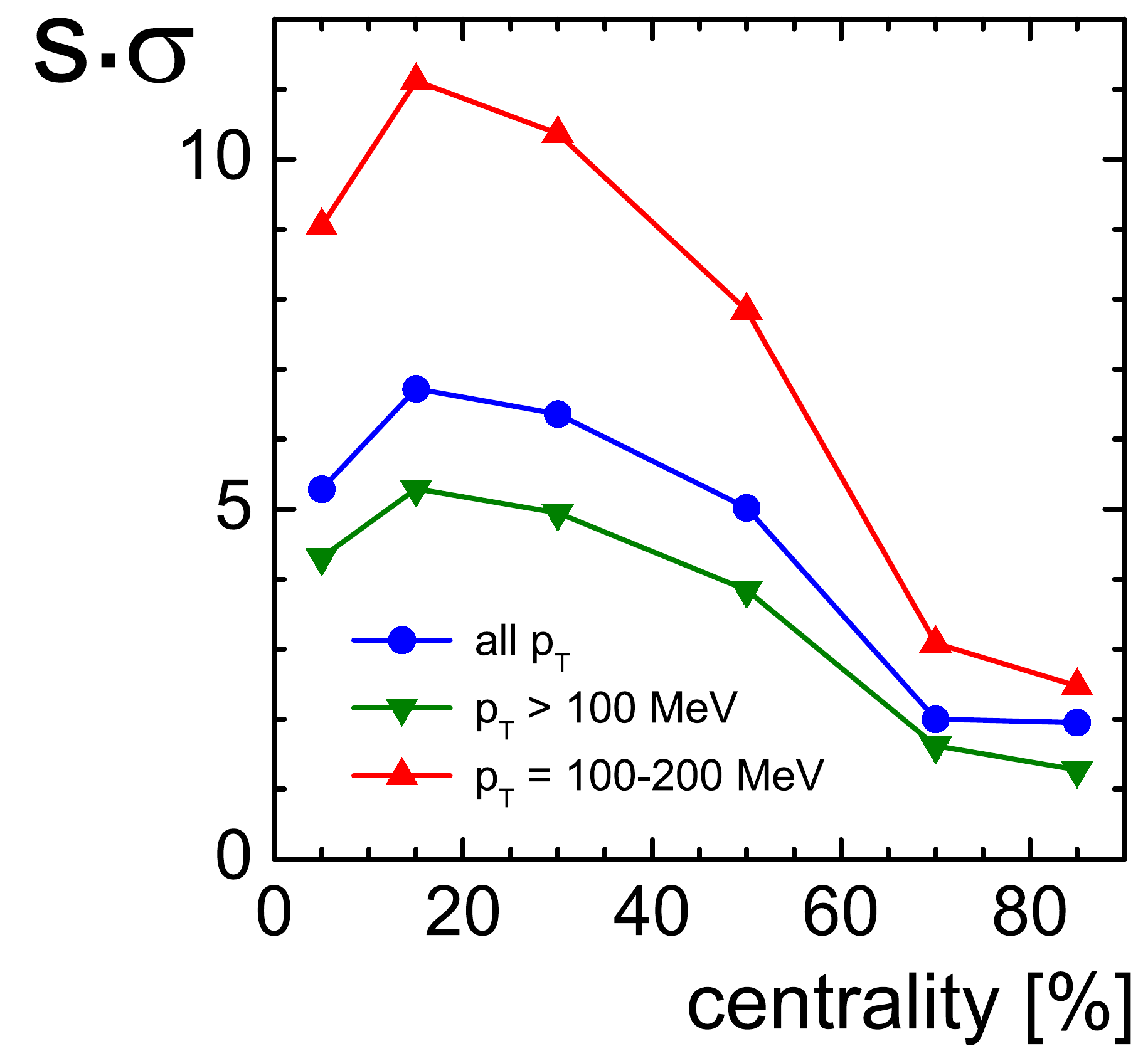}
 \caption{The same as Fig.~\ref{fig:W} for the normalized skewness.\label{fig:Ssig}}
\end{figure}
\begin{figure}[h!]
 \includegraphics[width=0.49\textwidth]{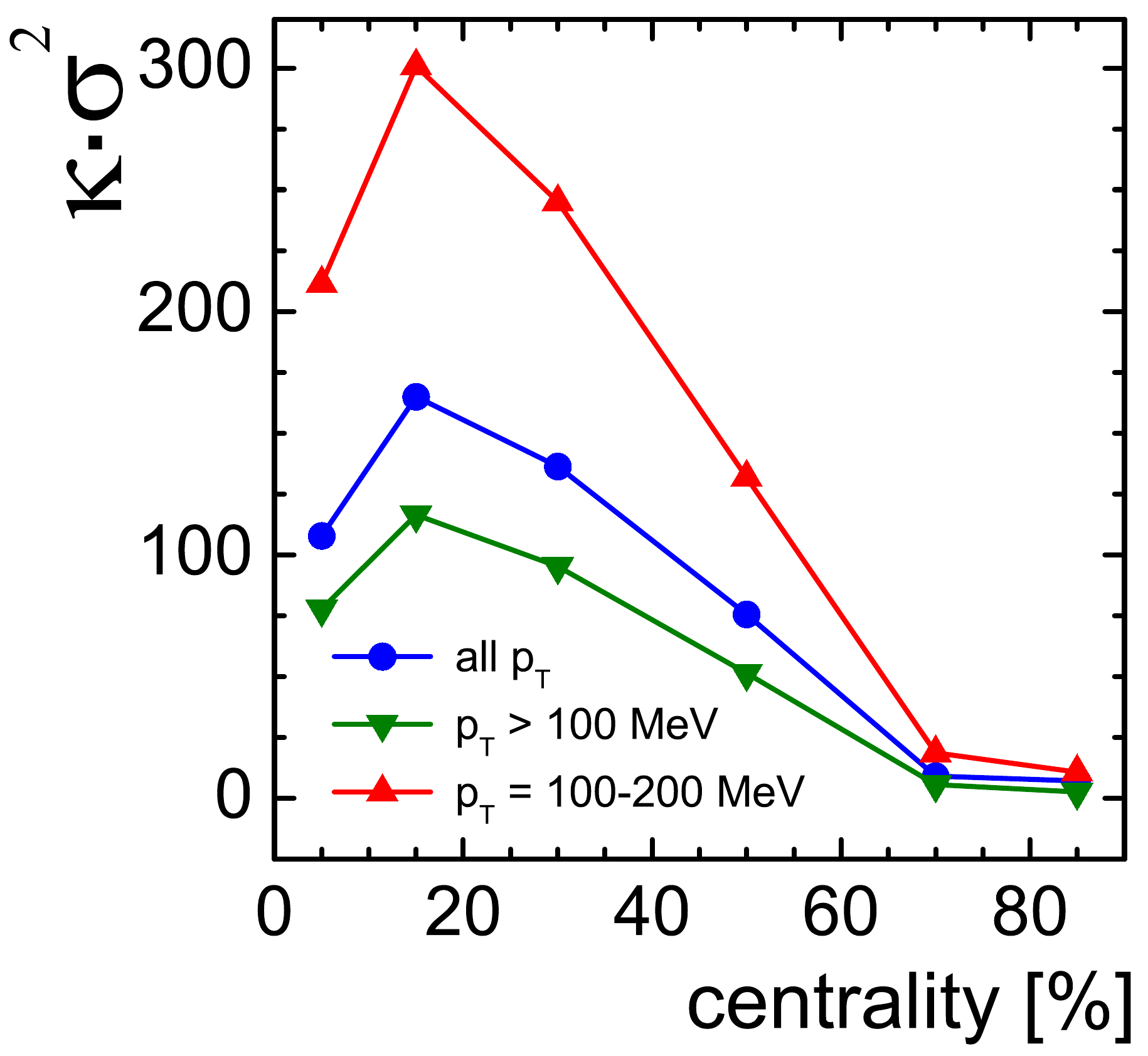}
 \caption{The same as Fig.~\ref{fig:Ssig} for the normalized kurtosis.\label{fig:ks2}}
\end{figure}

The results of the calculations using Eq.~(\ref{np}) are
substituted to Eqs.~(\ref{Np})-(\ref{ks2p}), then to
Eqs.~(\ref{N12})-(\ref{ks212}), and are presented in
Figs.~\ref{fig:W}-\ref{fig:ks2}. The error bars reflect the errors
in the $T$ and $\mu$ determination from the available experimental
data, see Fig.~\ref{fig:T-mu} (a), and are shown only for
$\omega$, see discussion below.
%
%
The scaled variances increase to some mild values, while the
normalized skewness and kurtosis are more sensitive variables. The
$p_T$ cut to $\Delta p_T=100-200$~MeV gives a factor of 2 increase
for the $S\cdot\sigma$ compared to other cases.
The normalized kurtosis reaches the values $\sim100$ even for the
measured $p_T$ range, while the $\Delta p_T=100-200$~MeV further
increases it three times to $\sim300$.
The scaling between the fluctuations according to Eq.~(\ref{wV})
holds for the $p_T$ cut $\Delta p_T=100-200$~MeV.

The error bars are about 30\%, 40\% and 70\% of the scaled
variance for the '$p_T>100$~MeV', the 'all $p_T$', and for the
'$\Delta p_T=100-200$~MeV' cases, correspondingly. They increase,
because of the increase of the unknown condensate part in the
corresponding cases. The error bars for the $S\cdot\sigma$ are of
the order of 100\% and even larger for the $\kappa\cdot\sigma^2$.
So large error bars mean that one can not predict an accurate
value of the fluctuations from the current data on mean
multiplicities. The experimental measurements of fluctuation may
show whether there is pion condensate or not.

Participant number (volume) fluctuation inside of a given
centrality is one of the most challenging ingredients of the
background. It is large for the scaled
variance~\cite{Begun:2012wq,Begun:2014boa}, and strongly increases
with the order of fluctuations measure~\cite{Begun:2016sop}.
Therefore, before making any conclusion out of the high order
fluctuations data, one should prove that participant number
fluctuations are under control.

The effect of cutting the $p_T$ range to $\Delta p_T=100-200$~MeV
gives much larger effect than measuring pions with all $p_T$. It
is an important advantage, because decreasing the $p_T$ requires
lower magnetic field and re-calibration of the
detectors~\cite{Shuryak:2014zxa}, while a $p_T$ cut can be implied
in the currently used software for the analysis of the events.

%
\section{Conclusions}\label{sect:Concl}
%

The normalized kurtosis is the most sensitive to chemical
non-equilibrium, pion condensation, and any other considered
effect.
It requires the largest number of measured events, and the
knowledge of the tails of the multiplicity distribution.
However, it rapidly grows if detector acceptance, size of the
system, or relative amount of particles in the condensate
increases.
It may allow to distinguish between equilibrium and
non-equilibrium models at the LHC.

The cut of the transverse momentum $p_T=100-200$~MeV for
positively (negatively) charged pions allows to increase the
relative amount of the condensate in the considered events, using
already measured data.
The possible increase of the normalized kurtosis is so large, that
one can check the intriguing possibility of high temperature
Bose-Einstein condensation of pions at the LHC experimentally.
%

\acknowledgments

The author thanks to M.I.~Gorenstein, M.~Chojnacki,
W.~Florkowski, Iu.~Karpenko, A.~Kisiel, M.~Mackowiak-Pawlowska and
L.~Tinti for fruitful comments and suggestions.
This work was supported by Polish National Science Center grant
No. DEC-2012/06/A/ST2/00390.

\bibliographystyle{h-physrev}
\bibliography{HMoments}{}

\end{document}